\documentclass[reprint, showpacs, superscriptaddress, aps, prl]{revtex4-2}

\usepackage{amsmath}
\usepackage{graphicx}
\usepackage{epstopdf}
\usepackage{color}
\usepackage{amssymb}
\usepackage{amsmath}
\usepackage{tabularx}
\usepackage{bm}
\usepackage{hyperref}
\usepackage{ltxtable}
\usepackage{natbib}
\usepackage{longtable}
\usepackage[normalem]{ulem}
\usepackage{float}

\newcommand{\be}{\begin{equation}}
\newcommand{\ee}{\end{equation}}
\newcommand{\bse}{\begin{subequations}}
\newcommand{\ese}{\end{subequations}}
\newcommand{\bea}{\begin{eqnarray}}
\newcommand{\eea}{\end{eqnarray}}
\newcommand{\ba}{\begin{array}}
\newcommand{\ea}{\end{array}}

\usepackage[usenames,dvipsnames]{xcolor}
\hypersetup{
colorlinks,
citecolor=blue,
filecolor=blue,
linkcolor=blue,
urlcolor=blue}

\begin{document}

\title{Classical 1/3 Nusselt number scaling in highly turbulent compressible convection}
\author{Harshit Tiwari}
\email{tharshit@iitk.ac.in}
\author{Mahendra K. Verma}
\email{mkv@iitk.ac.in}
\affiliation{Department of Physics, Indian Institute of Technology Kanpur, Kanpur 208016, India}

\date{\today}

\begin{abstract}
Planetary and stellar convection, which are compressible and turbulent, remain poorly understood. In this paper, we report numerical results on the scaling of Nusselt number ($\mathrm{Nu}$) and Reynolds number ($\mathrm{Re}$) for extreme convection. Using computationally-efficient MacCormack-TVD  finite difference method, we simulate compressible turbulent convection in a two-dimensional Cartesian box up to $\mathrm{Ra} = 10^{16}$, the highest $\mathrm{Ra}$ achieved so far, and in a three-dimensional box up to $\mathrm{Ra} = 10^{11}$. We show adiabatic temperature drop in the bulk flow, leading to the Reynolds number scaling $\mathrm{Ra}^{1/2}$. More significantly, we show classical $1/3$ Nusselt number scaling: $\mathrm{Nu} \propto \mathrm{Ra}^{0.32}$ in 2D, and $\mathrm{Nu} \propto \mathrm{Ra}^{0.31}$ in 3D up to the highest $\mathrm{Ra}$.
\end{abstract}

\maketitle

\onecolumngrid  
\vspace{-0.5cm} 
\begin{figure}
    \centering
    \includegraphics[width=\textwidth]{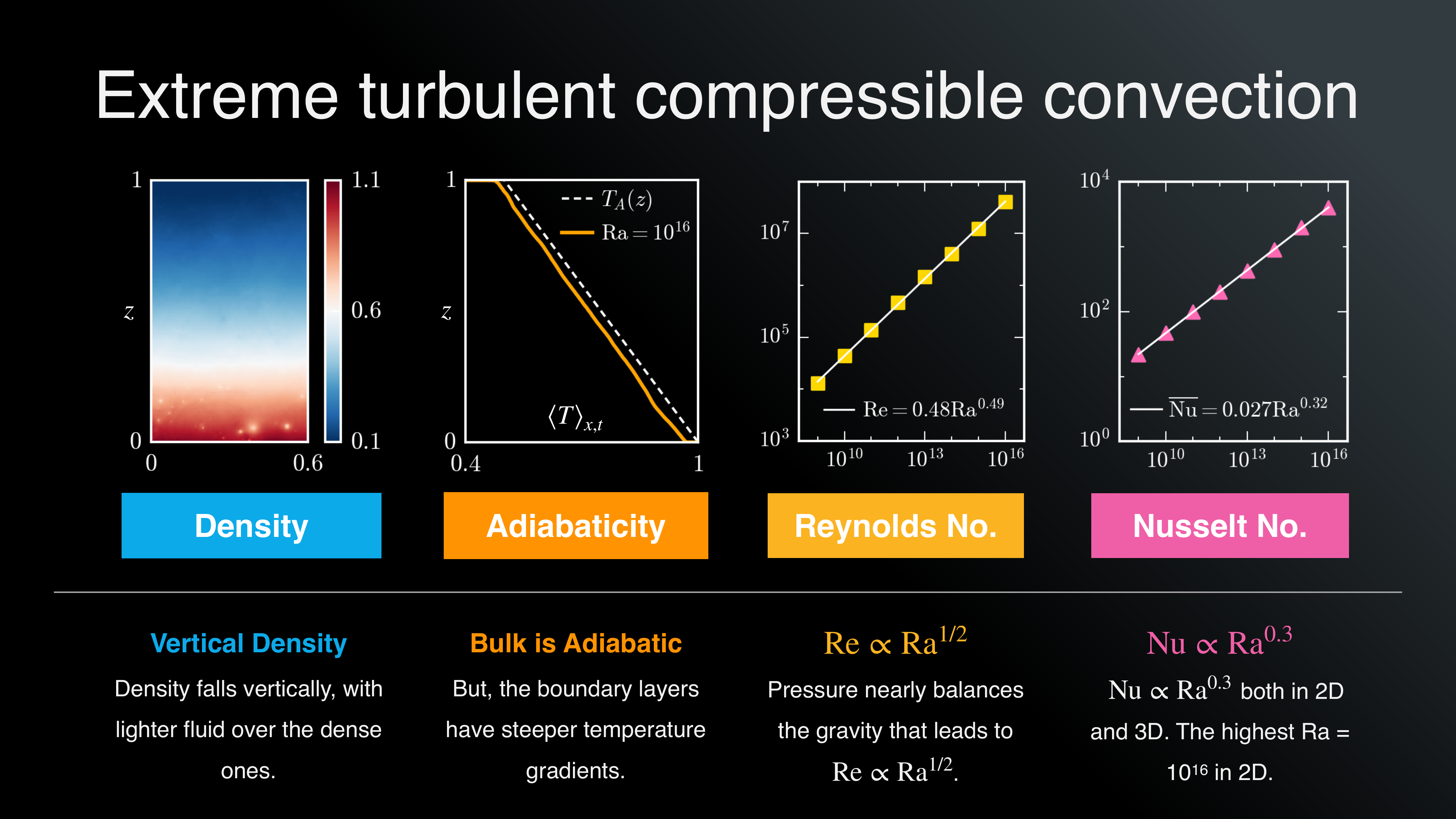}
    \label{fig:abstract}
\end{figure}
\twocolumngrid  

\onecolumngrid
\vspace{0.5cm}
\twocolumngrid


\section{Introduction}
Turbulent convection transports heat in planets, stars,  galaxies, tea kettles, boilers, etc. These systems belong to two categories: (a) Rayleigh-B\'{e}nard convection (RBC) satisfying Oberbeck-Boussinesq (OB) approximation, and (b) compressible convection~\cite{Siggia:ARFM1994,Lohse:ARFM2010,Ahlers:RMP2009,Chilla:EPJE2012,Verma:book:BDF,Schumacher:RMP2020,Spruit:ARAA1990}. Many terrestrial convection experiments and numerical simulations focus on category (a), where the relative changes in density $(\delta \rho)/\rho \ll 1$. For example, the thermal expansion coefficient ($\alpha$) for water at room temperature is approximately $2 \times 10^{-4} \mathrm{K}^{-1}$. Hence,  for the temperature difference $\Delta = 30$ Celsius between the  thermal plates, $(\delta \rho)/\rho \approx \alpha \Delta \approx 6 \times 10^{-3} $. On the contrary, astrophysical convection (e.g., solar convection) are compressible, for which ideal gas law is a reasonable assumption. For such systems, $\alpha = 1/T$, where $T$ is the temperature of the gas, and the temperature difference across the thermal boundaries is $O(T)$. Hence, $(\delta \rho)/\rho \approx 1$, making it necessary to employ compressible convection for astrophysical convection.  This paper focuses on the heat transport in compressible convection, which is not well studied.

RBC, which  is much better studied than compressible convection, is summarized below.  For RBC, the two nondimensional control parameters are Rayleigh number ($\mathrm{Ra}$), which measures the strength of buoyancy relative to the dissipation, and Prandtl number ($\mathrm{Pr}$), which is the ratio of kinematic viscosity and thermal diffusivity. The heat transport in the flow is quantified using a nondimensional parameter Nusselt number ($\mathrm{Nu}$).  Under the Boussinesq approximation, Shraiman and Siggia~\cite{Shraiman:PRA1990} derived exact relations between $\mathrm{Nu}$, $\mathrm{Ra}$, and $\mathrm{Pr}$.  For extreme turbulent regime, called the \textit{ultimate regime}, Kraichnan~\cite{Kraichnan:PF1962Convection} argued that $\mathrm{Nu}$ is proportional to $\mathrm{Ra}^{1/2}$; whereas Malkus~\cite{Malkus:PRSA1954} predicted that $\mathrm{Nu}$ is proportional to $\mathrm{Ra}^{1/3}$, which is called the \textit{classical scaling} for heat transport~\cite{Siggia:ARFM1994,Lohse:ARFM2010,Ahlers:RMP2009,Chilla:EPJE2012,Grossmann:JFM2000}.  Up to $\mathrm{Ra}=10^{12}$, the  $\mathrm{Nu}$ scaling exponent in experiments and numerical simulations is near $0.30$. For $\mathrm{Ra}$ beyond $10^{14}$, Chavanne et al.~\cite{Chavanne:PF2001}, He et al.~\cite{He:PRL2012}, and others show that the $\mathrm{Nu}$ scaling exponent increases gradually up to $0.38$ near $\mathrm{Ra}=10^{15}$. Based on these experiments and related numerical simulations~\cite{Zhu:PRL2018}, some researchers believe that the $\mathrm{Nu}$ scaling exponent will eventually reach $1/2$ at very high $\mathrm{Ra}$~\cite{Lohse:RMP2024,Shishkina:PRL2024}. But, based on other experiments~\cite{Niemela:Nature2000,Urban:PRL2012} and numerical simulations~\cite{Iyer:PNAS2020}, some researchers argue that the scaling $\mathrm{Nu} \sim \mathrm{Ra}^{1/3}$ will continue to be valid for all $\mathrm{Ra}$’s. In this paper, based on numerical simulations of  \textit{compressible convection}, we report that $\mathrm{Nu} \sim \mathrm{Ra}^{0.32}$ up to $\mathrm{Ra} = 10^{16}$ in two dimensions (2D), which is highest $\mathrm{Ra}$ achieved in numerical simulations, and $\mathrm{Nu} \sim \mathrm{Ra}^{0.31}$ up to $\mathrm{Ra} = 10^{11}$ in three dimensions (3D).

The ultimate regime is believed to be related to the nature of the boundary layer. Some researchers argue that near $\mathrm{Ra} \approx 10^{14}$, the viscous and thermal boundary layers become turbulent with logarithmic profiles that initiates a gradual transition to the ultimate regime~\cite{He:PRL2012, Zhu:PRL2018, Ahlers:PRL2012}. However, the emergence of turbulent boundary layer is still debated~\cite{Urban:PRL2011,Niemela:Nature2000,Urban:PRL2012}. In another front, Verma et al.~\cite{Verma:PRE2012} argued that the correlation between the temperature and vertical velocity is significant, and it alters the Nusselt number scaling from $\mathrm{Ra}^{1/2}$ to $\mathrm{Ra}^{1/3}$. In this paper, we observe similar correlations~\cite{Verma:PRE2012} in \textit{compressible convection}. 

Compared to RBC, there are much fewer works on compressible convection. John and Schumacher~\cite{John:PRF2023,John:JFM2023} performed numerical simulations of compressible convection simulation up to $\mathrm{Ra}=10^{6}$. Some other compressive simulations~\cite{Porter:AJS2000} employ inviscid flows whose $\mathrm{Ra}$ estimates are somewhat uncertain. Recently, Tiwari et al.~\cite{Tiwari:arxiv2024} simulated compressible convection up to $\mathrm{Ra}=10^{11}$ in 3D and up to $\mathrm{Ra}=10^{15}$ in 2D, and reported that $\mathrm{Nu} \sim \mathrm{Ra}^{0.3}$; however, their simulations are not fully resolved. In this paper, we extend the Ra of Tiwari et al.~\cite{Tiwari:arxiv2024} to $10^{16}$ in 2D, as well as increase the grid resolutions significantly.

For simulations, we extend the novel and stable numerical scheme proposed by Ouyang et al.~\cite{Ouyang:CG2013}, Yee~\cite{Yee:book}, and Liang et al.~\cite{Liang:IJNMF2007} for compressible hydrodynamic equations to turbulent convection; this exercise enabled us to perform convection simulations at extreme $\mathrm{Ra}$'s. For fluctuations, compressible simulations appear to be more stable than RBC simulations~\cite{Wesseling:book:CFD}. Hence, reaching very high $\mathrm{Ra}$ in compressible simulations seems to be easier than in RBC.

Three-dimensional turbulent simulations at high Rayleigh numbers are very expensive. Fortunately, the scaling of Reynolds and Nusselt numbers are similar in 2D and 3D~\cite{Pandey:PF2016}. Hence, we test the Nusselt number scaling for extreme $\mathrm{Ra}$ in 2D, but for moderate $\mathrm{Ra}$ in 3D.

\section*{METHODS}


\subsection*{Model}
\begin{figure}
	\centering
	\includegraphics[width=1\linewidth]{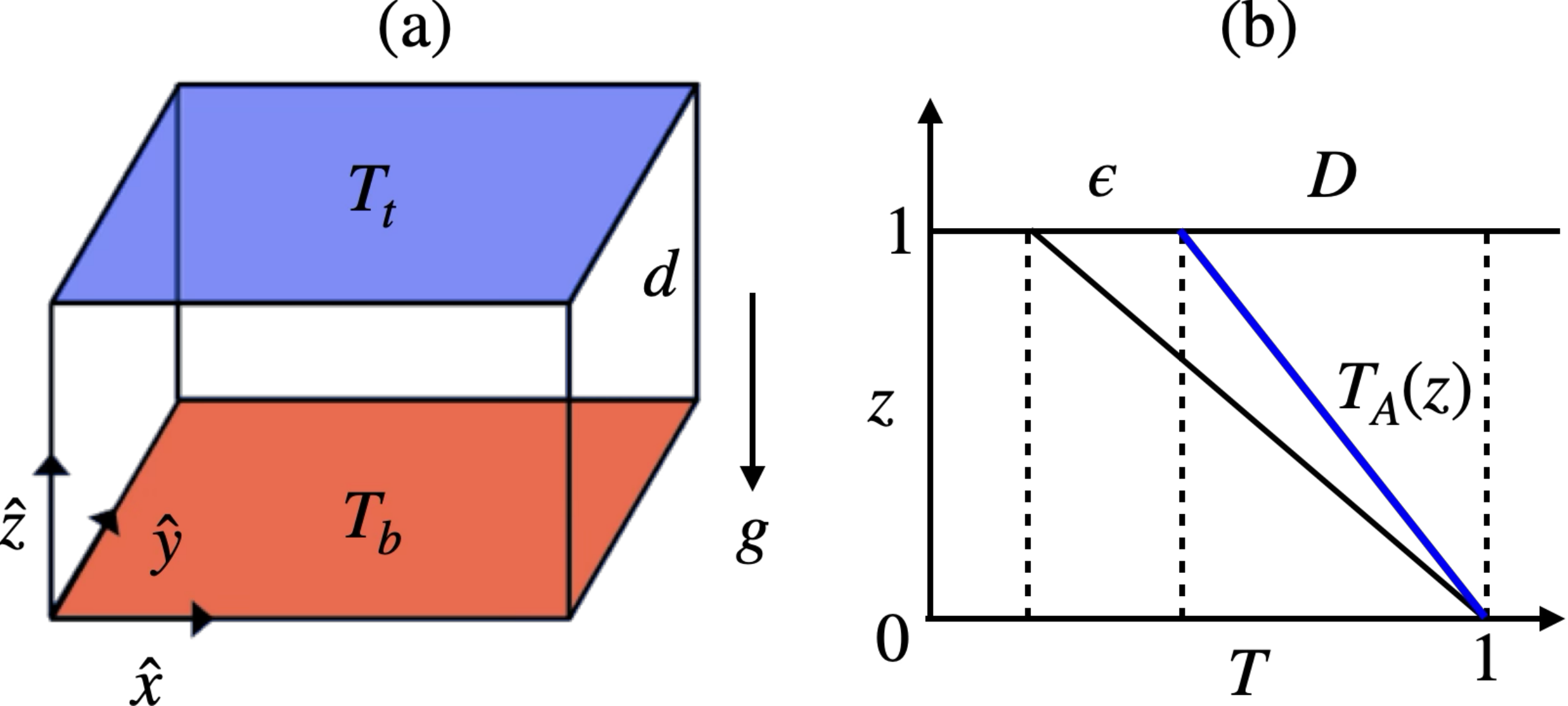}
	\caption{Illustration of the simulation box. (a) Flow inside a Cartesian box, which is heated at the bottom and cooled from the top with $T_b$ and $T_t$ temperatures, respectively. Gravity $-g\hat{z}$ is pointing downwards. (b) Non-dimensionalized adiabatic temperature profile (blue line) with $D$ temperature gradient and $\epsilon$ is the extra temperature drop at the top to start convection.}
	\label{fig:model}
\end{figure}

We simulate a fluid confined in a box of height $d$ with the bottom plate at temperature at $T_b$ and the top plate at $T_t$, with $\Delta=(T_b-T_t)$ (see Fig.~\ref{fig:model} (a)). The gravitational acceleration is $-g\hat{z}$. The boundary conditions are periodic horizontally, and no-slip at the top and bottom plates. We assume the fluid parameters---dynamic viscosity $\mu$, thermal conductivity $K$, and specific heat capacities at constant volume and pressure $C_p$ and $C_v$---to be constant everywhere. See Tiwari et al.~\cite{Tiwari:arxiv2024} for further details. 

As is customary, we simulate nondimensionalize flow equations. We use $d$ as the length scale, $T_b$ as the temperature scale, and the density at the bottom plate $\rho_b$ for the density scale. With this, the temperatures of the bottom and top plates are $1$ and $1-D-\epsilon$ respectively, where the dissipation number $D=gd/(C_p T_b)$ is the nondimensionalized adiabatic temperature gradient, and the superadiabaticity $\epsilon$ is the additional drop in nondimensional temperature at the top plate. See Fig.~\ref{fig:model} (b). The Prandtl number $\mathrm{Pr} = \nu/\kappa$, where $\nu = \mu/\rho_b$ and $\kappa= K/(C_p \rho_b)$ are respectively the kinematic viscosity and thermal diffusivity of the fluid, and Rayleigh number $\mathrm{Ra} =  \epsilon g d^3/(\nu \kappa)$, where the factor $\epsilon$ provides correction due to the adiabatic temperature drop. 

The non-dimensionalized governing equations in conservative form are~\cite{Spiegel:AJ1965, Graham:JFM1975}:
\begin{equation}
     \frac{\partial \rho}{\partial t} + \frac{\partial}{\partial x_i}(\rho u_i) = 0,\label{eq:continuity_nondim}
\end{equation}
\begin{equation}
    \frac{\partial}{\partial t}(\rho u_i) + \frac{\partial}{\partial x_j}(\rho u_i u_j + \delta_{ij}p - \tau_{ij}) = -\frac{1}{\epsilon}\rho\delta_{iz},\label{eq:momentum_nondim}
\end{equation}
\begin{equation}
    \frac{\partial E}{\partial t} + \frac{\partial}{\partial x_i} \left( u_i(E + p) - \frac{1}{\epsilon D \sqrt{\mathrm{Ra} Pr}} \frac{\partial T}{\partial x_i} - u_j \tau_{ij} \right) = 0,\label{eq:energy_nondim}
\end{equation}
where, $u_i$ and $T$ are the non-dimensionalized velocity and temperature fields respectively.   The non-dimensionalized pressure for the ideal gas is
\begin{equation}
    p=(\gamma-1) \frac{\rho T}{\gamma \epsilon D},
\end{equation}
 where polytropic index $\gamma=C_p/C_v$. The stress tensor is 
\begin{equation}
    \tau_{ij} = \sqrt{\frac{\mathrm{Pr}}{\mathrm{Ra}}} \left( \partial_j u_i + \partial_i u_j + \frac{2}{3} \partial_m u_m \delta_{ij} \right),
\end{equation}
and the total energy density is
\begin{equation}
    E=\rho(u^2/2 + T/(\gamma \epsilon D) + z/\epsilon).
\end{equation}
The horizontal and temporal averaged kinetic energy dissipation rate at height $z$ is $\zeta(z)= \langle S_{ij} \tau_{ij} \rangle/\langle \rho \rangle$, where $S_{ij}=(\delta_i u_j+\delta_j u_i)/2$ is the strain rate tensor. In all our simulations, $\mathrm{min}( \eta(z)/\Delta z)$ and  $\mathrm{min}( \eta(z)/\Delta x)$ are greater than 0.5, where $\eta(z) = (\nu^3 /  \zeta(z) )^{1/4}$ is the Kolmogorov length~\cite{Jagannathan:JFM2016,John:PRF2023} (see Table ~\ref{table}). This ensures that the simulations are fully resolved.



\subsection*{Numerical Scheme}

To solve the partial differential equations~\ref{eq:continuity_nondim},~\ref{eq:momentum_nondim},~\ref{eq:energy_nondim}, we employ MacCormack-TVD (total variation diminishing) finite-difference scheme on collocated grid~\cite{Wesseling:book:CFD,Ouyang:CG2013}. The
MacCormack-TVD scheme is second-order accurate in space and time~\cite{Wesseling:book:CFD}. We employ non-uniform tangent-hyperbolic grid in the $z$ direction to increase resolution near the boundaries. The grids in the $x$ and $y$ directions are uniform. Boundary points are computed using second-order forward differences at the bottom plate and backward differences at the top plate.

The generalized equation is written in the following conservative form:
\begin{equation}
    \frac{\partial Q}{\partial t} + \frac{\partial F_i}{\partial x_i} = S_i,\label{eq:vector_eqn}
\end{equation}
where $Q$ is a scalar field, $F_i$ and $S_i$ are the flux and the source term in $i$-direction. By employing the operator-splitting method, this equation is decomposed into three one-dimensional equations and solved using the MacCormack scheme, which takes the predictor-corrector average for the next time step. See Tiwari et al.~\cite{Tiwari:arxiv2024} for details. The TVD correction term is added to preserve monotonicity and prevent spurious oscillations in the solution~\cite{Ouyang:CG2013}. The MacCormack scheme, combined with TVD correction, is particularly advantageous for compressible convection problems due to its simplicity, second-order accuracy, and ability to handle shocks and steep gradients effectively.

\section*{RESULTS}

\subsection*{Numerical simulations}

We simulate compressible fluid inside a Cartesian box using computationally-efficient MacCormack TVD (total variation diminishing) finite difference method~\cite{Wesseling:book:CFD,Ouyang:CG2013} (see Methods). The aspect ratio $\Gamma$ of 2D and 3D boxes are given in Table~\ref{table}. Our simulations are well resolved, both in the bulk and in the boundary layers~\cite{Jagannathan:JFM2016,John:PRF2023}. We perform our parallel runs on several GPUs that enables high resolution simulations in a reasonable time. For instance, our largest 3D simulation on $513^2 \times 2049$ grid ran for 7 days on 2 nodes of Polaris (with four A100 GPUs each) at Argonne National Laboratory, while the largest 2D run on $24001^2$ grid ran for 5 days using 8 nodes.

\begin{figure}%
	\centering
	\includegraphics[width=1\linewidth]{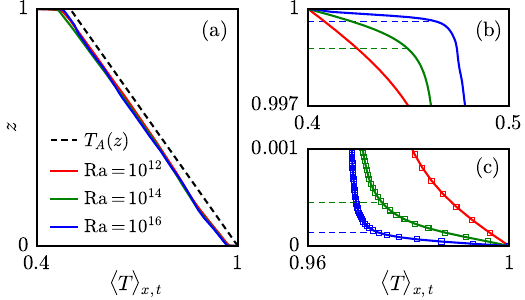}
	\caption{For 2D simulations with $\mathrm{Ra} = 10^{12}$, $ 10^{14}$,   and $10^{16}$, profiles of horizontally and temporally averaged $\langle T \rangle_{x,t}(z)$. (a) Plots of  $\langle T \rangle_{x,t}(z)$ for $\mathrm{Ra} = 10^{12}$ (red), $ 10^{14}$ (green), and $ 10^{16}$ (blue).  $T_A(z)$ (black dashed line) represents the adiabatic temperature profile. (b, c) Zoomed-in views near the top and bottom boundaries.}
	\label{fig:T_z_profile}
\end{figure}

\begin{figure*}%
	\centering
	\includegraphics[width=1\textwidth]{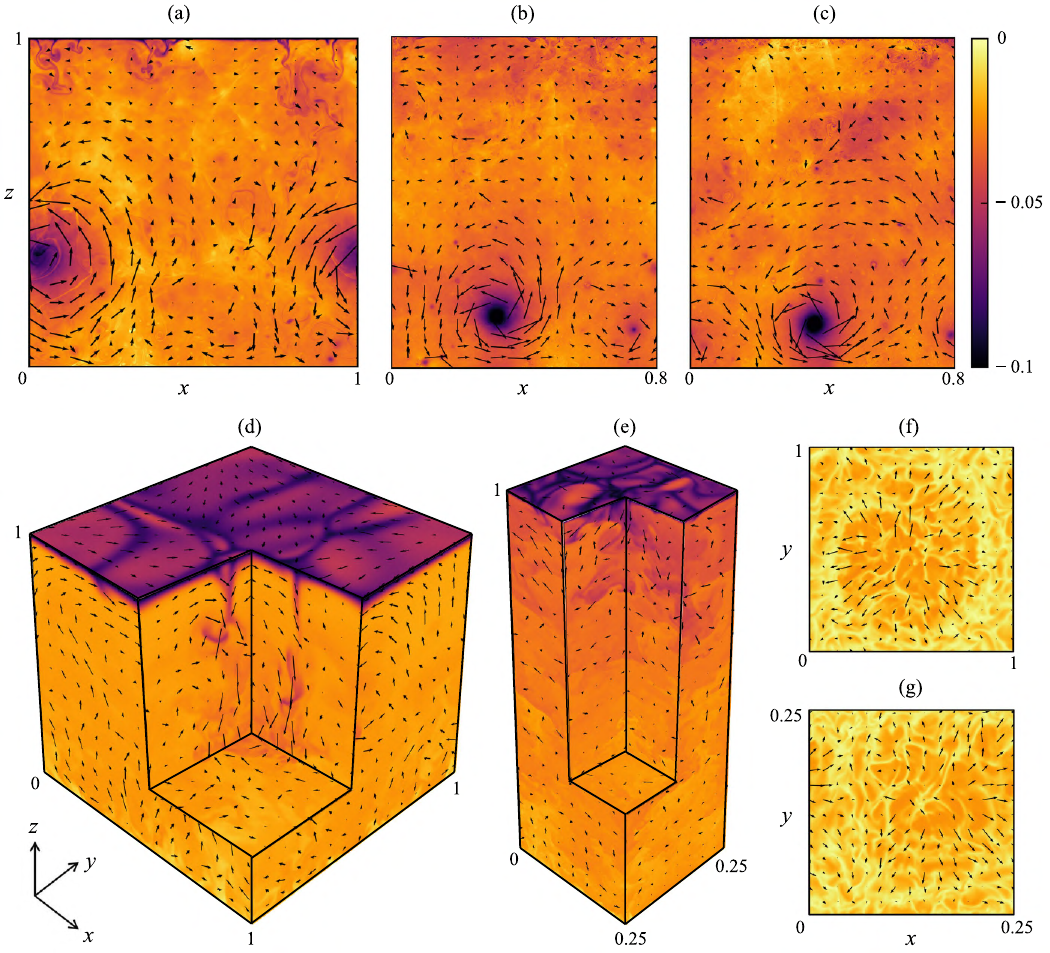}
	\caption{ Field profiles with  velocity vector plot ($\bf{{u}}$) and superadiabatic temperature density plot ($\theta(\textbf{r})=T(\textbf{r})-T_A (z)$) for 2D and 3D convection:  For 2D, (a) $\mathrm{Ra}=10^{12}$, (b) $10^{14}$, and (c) $10^{16}$; and for 3D, (d) $\mathrm{Ra}=10^{9}$ and (e) $10^{11}$. (f, g)  Field profiles near the bottom plate for (d) and (e). }
	\label{fig:field_profiles}
\end{figure*}

\begin{table*}
    \footnotesize
	\centering
	\caption{For our 2D and 3D runs with polytropic index $\gamma=1.3$, Prandtl number $\mathrm{Pr} = 0.7$, dissipation number $D = 0.5,$ and superadiabaticity $\epsilon=0.1$:  Rayleigh number $\mathrm{Ra}$,  grid size,  aspect ratio $\Gamma$,  Reynolds number $\mathrm{Re}$,  mean Nusselt number based on boundary layers ($\overline{\mathrm{Nu}}$), the number of grid points in the top and bottom boundary layers ($N_t^{\mathrm{BL}}$, $N_b^{\mathrm{BL}}$), the thicknesses of top and bottom thermal boundary layers ($\lambda_t$, $\lambda_b$), $\delta_{\bot} = \mathrm{min}(\eta(z)/\Delta x)$, and $\delta_{\parallel}= \mathrm{min}( \eta(z)/\Delta z)$. Here, $\eta(z)$ is the Kolmogorov length at height $z$.   }
	\vspace{5pt}
	\setlength{\tabcolsep}{7.5pt} 
	\renewcommand{\arraystretch}{1.2} 
	\begin{tabular}{|l|c|c|c|c|c|c|c|c|c|c|c|}
		\hline
		Run & $\mathrm{Ra}$ & Grid Size & $\Gamma$  & $\mathrm{Re}$ & $\overline{\mathrm{Nu}}$ & $N_t^{\mathrm{BL}}$ & $N_b^{\mathrm{BL}}$ & $\lambda_t$ & $\lambda_b$ & $\delta_{\bot}$ & $\delta_{\parallel}$\\
		\hline
		$1$ & $10^{9}$ & $513^2$ & $1$ & $(1.3 \pm 0.1) \times 10^{4}$ & $22 \pm 2$ & $32$ & $11$ & $0.038$ & $0.012$ & $0.82$ & $1.08$\\
		$2$ & $10^{10}$ & $1025^2$ & $1$ & $(4.4 \pm 0.2) \times 10^{4}$ & $47 \pm 3$ & $32$ & $11$ & $0.0180$ & $0.0061$ & $0.66$ & $0.88$\\
		$3$ & $10^{11}$ & $2049^2$ & $1$ & $(1.38 \pm 0.08) \times 10^{5}$ & $103 \pm 12$ & $30$ & $10$ & $0.0084$ & $0.0029$ & $0.57$ & $0.84$\\
		$4$ & $10^{12}$ & $4097^2$ & $1$ & $(4.7 \pm 0.2) \times 10^{5}$ & $221 \pm 23$ & $31$ & $10$ & $0.0036$ & $0.0013$ & $0.57$ & $0.89$\\
		$5$ & $10^{13}$ & $8193^2$ & $1$ & $(1.46 \pm 0.02) \times 10^{6}$ & $430 \pm 35$ & $34$ & $10$ & $0.00235$ & $0.00066$ & $0.56$ & $0.84$\\
		$6$ & $10^{14}$ & $12001^2$ & $0.8$ & $(4.04 \pm 0.05) \times 10^{6}$ & $895 \pm 112$ & $28$ & $9$ & $0.00123$ & $0.00046$ & $0.51$ & $0.57$\\
		$7$ & $10^{15}$ & $16385^2$ & $0.8$ & $(1.24 \pm 0.02) \times 10^{7}$ & $2130 \pm 220$ & $25$ & $8$ & $0.00067$ & $0.00026$ & $0.52$ & $0.59$\\
		$8$ & $10^{16}$ & $24001^2$ & $0.8$ & $(4.1 \pm 0.1) \times 10^{7}$ & $4120 \pm 510$ & $23$ & $7$ & $0.00039$ & $0.00014$ & $0.51$ & $0.55$\\
		\hline
		$9$ & $10^{7}$ & $129^3$ &  $1$ &$(7.1 \pm 0.3) \times 10^{2}$ & $8.2 \pm 0.4$ & $25$ & $8$ & $0.11$ & $0.04$ & $0.76$ & $0.95$\\
		$10$ & $10^{8}$ & $257^3$ &  $1$ &$(2.1 \pm 0.1) \times 10^{3}$ & $14.3 \pm 0.7$ & $33$ & $12$ & $0.054$ & $0.016$ & $0.76$ & $0.95$\\
		$11$ & $10^{9}$ & $513^3$ &  $1$ &$(6.0 \pm 0.1) \times 10^{3}$ & $27.5 \pm 0.7$ & $32$ & $11$ & $0.029$ & $0.008$ & $0.72$ & $0.89$\\
		$12$ & $10^{7}$ & $33^2 \times 129$ &  $0.25$ &$(3.6 \pm 0.5) \times 10^{2}$ & $5 \pm 1$ & $28$ & $9$ & $0.14$ & $0.04$ & $0.76$ & $0.95$\\
		$13$ & $10^{8}$ & $65^2 \times 257$ &  $0.25$ &$(1.5 \pm 0.1) \times 10^{3}$ & $14 \pm 1$ & $32$ & $12$ & $0.049$ & $0.015$ & $0.76$ & $0.95$\\
		$14$ & $10^{9}$ & $129^2 \times 513$ &  $0.25$ &$(4.3 \pm 0.3) \times 10^{3}$ & $27 \pm 2$ & $31$ & $11$ & $0.028$ & $0.009$ & $0.72$ & $0.89$\\
		$15$ & $10^{10}$ & $257^2 \times 1025$ & $0.25$ & $(1.22 \pm 0.08) \times 10^{4}$ & $56 \pm 3$ & $30$ & $10$ & $0.0141$ & $0.0052$ & $0.69$ & $0.84$\\
		$16$ & $10^{11}$ & $513^2 \times 2049$ & $0.25$ & $(3.82 \pm 0.06) \times 10^{4}$ & $120 \pm 4$ & $29$ & $9$ & $0.0074$ & $0.0024$ & $0.66$ & $0.81$\\
		\hline
	\end{tabular}
	\label{table}
\end{table*}

For our numerical simulations, we fix the polytropic index $\gamma=1.3$, the dissipation number $D = 0.5$, the superadiabaticity $\epsilon = 0.1$, $\mathrm{Pr} = 0.7$, and vary $\mathrm{Ra}$ from $10^9$ to $10^{16}$ for 2D, and from $10^7$ to $10^{11}$ for 3D (see Methods). The parameters are listed in Table~\ref{table}. The highest grid resolution for 2D convection is $24001^2$, and for 3D convection is $513^2 \times 2049$. In all our simulations, the number of vertical grid points in the boundary layers are more than 6, thus satisfying \textit{Grötzbach resolution criterion}~\cite{Grotzbach:JCP1983}. More importantly, $\mathrm{min}( \eta(z)/\Delta z)$ and  $\mathrm{min}( \eta(z)/\Delta x)$ are greater than 0.5, where $\eta(z) = (\nu^3 /  \zeta(z) )^{1/4}$ is the Kolmogorov length, and   $ \zeta(z) $ is the  kinetic energy dissipation rate at height $z$~\cite{Jagannathan:JFM2016,John:PRF2023} (see Table ~\ref{table} and Methods). Hence, our simulations are well resolved.

Compressible convection starts when the vertical temperature gradient becomes steeper than adiabatic temperature gradient $g/C_p$~\cite{Spiegel:AJ1965}. This is called \textit{Schwarzschild criterion}. Surprisingly, the adiabatic temperature gradient $g/C_p$ is maintained in the bulk for all $\mathrm{Ra}$'s studied in our work (Fig.~\ref{fig:T_z_profile}). However, the temperature gradients in the top and bottom boundary layers are steeper than $g/C_p$, which leads to strong convection in these regions. As shown in Fig.~\ref{fig:T_z_profile}, the vertical temperature (averaged horizontally) drops linearly, similar to the adiabatic cooling. This is in contrast to RBC where the average bulk temperature is nearly constant. 

In Fig.~\ref{fig:field_profiles} we illustrate the velocity field and the superadiabatic temperature field $\theta({\bf r})=T({\bf r})-T_A (z)$, which is the difference between the temperature and adiabatic profile, for $\mathrm{Ra}$ = $10^{12}$, $10^{14}$, and $10^{16}$ in 2D, and for $\mathrm{Ra}$ = $10^9$ and $10^{11}$ in 3D. The flow structures become thinner with the increase of $\mathrm{Ra}$. In addition, the density at the bottom is larger than that at the top, which is opposite to that in RBC.  Also, the thermal boundary layer at the top is thicker than that at the bottom, consistent with the earlier observations~\cite{John:JFM2023} (see Fig.~\ref{fig:T_z_profile}). In contrast, the top and boundary layers are nearly symmetric in RBC. A \href{https://www.youtube.com/watch?v=mypHAJt8-VA}{movie} for $\mathrm{Ra}$ = $10^{12}$ illustrates the dynamical evolution of the flow.

\subsection*{Reynolds Number Scaling}

\begin{figure}
\centering
\includegraphics[width=0.93\linewidth]{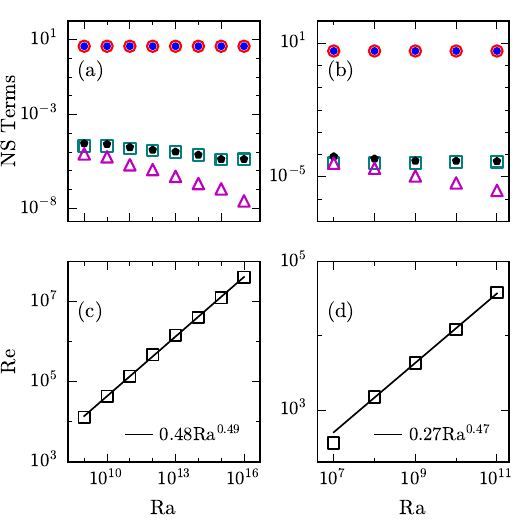}
\caption{Reynolds number scaling: For 2D (a) and 3D (b), plots of the averages of various terms of the momentum equation: $-g \langle \rho \rangle$ (blue dots), $\langle \partial_z p \rangle$ (red unfilled circles), $\langle \partial_i(\rho u_i u_z) \rangle$ (teal squares), viscous term (magenta triangles), and $- \langle \rho g \rangle - \langle \partial_z p \rangle$ (black pentagon). Note that $- \langle \rho g \rangle \approx \langle \partial_z p \rangle$, and $- \langle \rho g \rangle - \langle \partial_z p \rangle \approx \langle \partial_i(\rho u_i u_z) \rangle$. The latter balance equation yields the Reynolds number scaling, $\mathrm{Re} \propto \mathrm{Ra}^{1/2}$, which is borne out in 2D (c) and 3D (d).}
\label{fig:reynolds}
\end{figure}

We compare the four terms of the momentum equation for various $\mathrm{Ra}$’s. Their volume averages are plotted in Fig.~\ref{fig:reynolds} (a,b) for 2D and 3D respectively. For all the runs, the pressure gradient $\langle \partial_z p \rangle$ nearly balances the gravitational term $-\langle \rho g \rangle$. In addition, the viscous term is negligible, and the small difference $- \langle \rho g \rangle - \langle \partial_z p \rangle$ nearly equals the nonlinear term, that is, $- \langle \rho g \rangle - \langle \partial_z p \rangle \approx \langle  \partial_i(\rho u_i u_z) \rangle$. Based on the above estimates, we derive that the Reynolds number $\mathrm{Re} = Ud/\nu \sim (\mathrm{Ra}/\mathrm{Pr})^{1/2}$.  Using our numerical data we verify the balances in the momentum equation and the $\mathrm{Re}$ scaling (see Fig.~\ref{fig:reynolds} (c,d)). We remark that the relations between various terms of the respective RBC equation are more complex~\cite{Siggia:ARFM1994, Ahlers:RMP2009, Grossmann:JFM2000, Verma:book:BDF, Landau:book:Fluid}. 

The above observations shed important light on the dynamics of compressible convection. The balance, $\langle \partial_z p \rangle \approx -\langle \rho g \rangle$
at the onset of convection leads to the Schwarzschild condition~\cite{Landau:book:Fluid}.  Interestingly, the above balance also holds for large $\mathrm{Ra}$'s. This is because the internal energy ($\rho C_v T \sim p$) dominates the fluid kinetic energy ($(1/2) \rho u^2$) even in the turbulent regime. For example, the average Mach number for the 2D flow with $\mathrm{Ra}$ = $10^{16}$ is $0.2$. We will detail these results in a subsequent paper. 


\subsection*{Nusselt Number Scaling}

\begin{figure}
	\centering
	\includegraphics[width=1\linewidth]{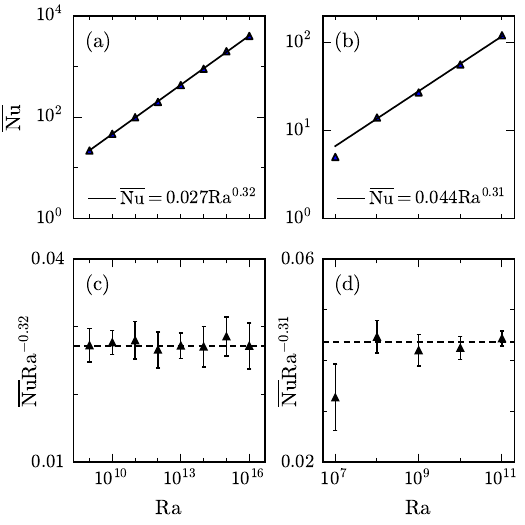}
	\caption{Nusselt number scaling  for  (a) 2D and (b) 3D.  The averaged boundary-layer based $\mathrm{\overline{Nu}}$ exhibit near classical scaling.  (c,d) exhibit the scaled $\overline{\mathrm{Nu}}\mathrm{Ra}^\alpha$ plots. }
	\label{fig:nusselt}
\end{figure}

In compressible turbulent convection, the Nusselt number $\mathrm{Nu}$ computed near the bottom plate ($z=0$) and top plate ($z=1$) is given by\cite{Verhoeven:AJ2015, John:JFM2023} 
\begin{equation}
    \mathrm{Nu}_{z=0,1} = -\frac{1}{\epsilon} \left. \frac{d \langle \theta \rangle_{A,t}}{dz} \right\vert_{z =0,1}. \label{eq:Nu}
\end{equation}
Here $\langle . \rangle_{A,t}$ represents the horizontal and temporal averages. Consequently, the mean Nusselt number at the boundaries is given by $\overline{\mathrm{Nu}} = (\mathrm{Nu}_{z=0} + \mathrm{Nu}_{z=1})/2$.
Using the steady-state numerical data, we compute the averaged Nusselt number $\mathrm{\overline{Nu}}$ at the boundaries (see Table ~\ref{table}). We plot $\mathrm{\overline{Nu}}$ for various $\mathrm{Ra}$'s in Fig.~\ref{fig:nusselt}. For the available data,   $\mathrm{\overline{Nu}} = (0.027 \pm 0.001) \mathrm{Ra}^{(0.32 \pm 0.002)}$ for 2D up to $\mathrm{Ra} = 10^{16}$ (largest $\mathrm{Ra}$ achieved so far), and $\mathrm{\overline{Nu}} = (0.044 \pm 0.007) \mathrm{Ra}^{(0.31 \pm 0.007)}$ for 3D up to $\mathrm{Ra} = 10^{11}$. Although the ultimate regime is proposed for RBC, we have tested its existence in compressible convection and demonstrate that compressible convection exhibits classical $1/3$ scaling rather than ultimate $1/2$ scaling.

\section*{DISCUSSION}

In this paper, we performed fully-resolved compressible convection simulations, extending up to $\mathrm{Ra} = 10^{16}$ in 2D and $\mathrm{Ra} = 10^{11}$ in 3D. These high-resolution simulations ensure accurate representation of both small- and large-scale features of compressible turbulent convection, allowing us to capture the critical dynamics and scaling laws  reliably. In our simulations, the internal energy dominates the fluid kinetic energy that leads to an adiabatic temperature drop in the bulk for all $\mathrm{Ra}$'s. Based on these observations, we show that Reynolds number $\mathrm{Re} \propto \mathrm{Ra}^{1/2}$. In spite of several critical differences between compressible convection and RBC, we observe classical Nusselt number scaling: $\mathrm{\overline{Nu}}  \propto \mathrm{Ra}^{0.32}$ in 2D for $\mathrm{Ra}$ up to $10^{16}$, and $\mathrm{\overline{Nu}}   \propto \mathrm{Ra}^{0.31}$ in 3D for $\mathrm{Ra}$ up to $10^{11}$.  These results are of major importance for modelling heat transport in planetary and stellar atmospheres.

Astrophysical fluids are compressible; hence,  compressible convection is more suitable than RBC for such flows. In particular, compressible convection will better model the stellar and terrestrial atmospheres than RBC.  The convective zone of   the Sun is compressive with $\mathrm{Ra}$ near $10^{24}$ and thermal Prandtl number near $10^{-7}$~\cite{Schumacher:RMP2020}. In addition, solar convection is coupled to the magnetic field~\cite{Parker:ARAA1970}. In comparison to the solar convection, our numerical simulations lack magnetic field and have moderate Prandtl numbers. Yet, the convection profiles of our simulation (e.g., adiabatic profile) are in reasonable agreement with those in the Sun~\cite{Miesch:LRSP2005}.   In future, we plan to simulate turbulent magnetoconvection with higher $\mathrm{Ra}$ and lower $\mathrm{Pr}$ values and test the $\mathrm{Nu}$ and $\mathrm{Re}$ scaling.

\section*{Acknowledgment}

\begin{acknowledgments}
We thank K. R. Sreenivasan, Jörg Schumacher, Lekha Sharma, and Dhananjay Singh for useful discussions. We are thankful to Argonne Leadership Computing Facility (ALCF) for providing us computer time through Director’s Discretionary Program. Part of the simulations were performed on Param Sanganak of IIT Kanpur.
\end{acknowledgments}

\bibliography{journal, book, arxiv_book, arxiv_journal}

\begin{thebibliography}{39}%
\makeatletter
\providecommand \@ifxundefined [1]{%
 \@ifx{#1\undefined}
}%
\providecommand \@ifnum [1]{%
 \ifnum #1\expandafter \@firstoftwo
 \else \expandafter \@secondoftwo
 \fi
}%
\providecommand \@ifx [1]{%
 \ifx #1\expandafter \@firstoftwo
 \else \expandafter \@secondoftwo
 \fi
}%
\providecommand \natexlab [1]{#1}%
\providecommand \enquote  [1]{``#1''}%
\providecommand \bibnamefont  [1]{#1}%
\providecommand \bibfnamefont [1]{#1}%
\providecommand \citenamefont [1]{#1}%
\providecommand \href@noop [0]{\@secondoftwo}%
\providecommand \href [0]{\begingroup \@sanitize@url \@href}%
\providecommand \@href[1]{\@@startlink{#1}\@@href}%
\providecommand \@@href[1]{\endgroup#1\@@endlink}%
\providecommand \@sanitize@url [0]{\catcode `\\12\catcode `\$12\catcode `\&12\catcode `\#12\catcode `\^12\catcode `\_12\catcode `\%12\relax}%
\providecommand \@@startlink[1]{}%
\providecommand \@@endlink[0]{}%
\providecommand \url  [0]{\begingroup\@sanitize@url \@url }%
\providecommand \@url [1]{\endgroup\@href {#1}{\urlprefix }}%
\providecommand \urlprefix  [0]{URL }%
\providecommand \Eprint [0]{\href }%
\providecommand \doibase [0]{https://doi.org/}%
\providecommand \selectlanguage [0]{\@gobble}%
\providecommand \bibinfo  [0]{\@secondoftwo}%
\providecommand \bibfield  [0]{\@secondoftwo}%
\providecommand \translation [1]{[#1]}%
\providecommand \BibitemOpen [0]{}%
\providecommand \bibitemStop [0]{}%
\providecommand \bibitemNoStop [0]{.\EOS\space}%
\providecommand \EOS [0]{\spacefactor3000\relax}%
\providecommand \BibitemShut  [1]{\csname bibitem#1\endcsname}%
\let\auto@bib@innerbib\@empty
\bibitem [{\citenamefont {Siggia}(1994)}]{Siggia:ARFM1994}%
  \BibitemOpen
  \bibfield  {author} {\bibinfo {author} {\bibfnamefont {E.~D.}\ \bibnamefont {Siggia}},\ }\bibfield  {title} {\bibinfo {title} {{High Rayleigh number convection}},\ }\href@noop {} {\bibfield  {journal} {\bibinfo  {journal} {Annu. Rev. Fluid Mech.}\ }\textbf {\bibinfo {volume} {26}},\ \bibinfo {pages} {137} (\bibinfo {year} {1994})}\BibitemShut {NoStop}%
\bibitem [{\citenamefont {Lohse}\ and\ \citenamefont {Xia}(2010)}]{Lohse:ARFM2010}%
  \BibitemOpen
  \bibfield  {author} {\bibinfo {author} {\bibfnamefont {D.}~\bibnamefont {Lohse}}\ and\ \bibinfo {author} {\bibfnamefont {K.-Q.}\ \bibnamefont {Xia}},\ }\bibfield  {title} {\bibinfo {title} {{Small-scale properties of turbulent Rayleigh{\textendash}B{\'e}nard convection}},\ }\href@noop {} {\bibfield  {journal} {\bibinfo  {journal} {Annu. Rev. Fluid Mech.}\ }\textbf {\bibinfo {volume} {42}},\ \bibinfo {pages} {335} (\bibinfo {year} {2010})}\BibitemShut {NoStop}%
\bibitem [{\citenamefont {Ahlers}\ \emph {et~al.}(2009)\citenamefont {Ahlers}, \citenamefont {Grossmann},\ and\ \citenamefont {Lohse}}]{Ahlers:RMP2009}%
  \BibitemOpen
  \bibfield  {author} {\bibinfo {author} {\bibfnamefont {G.}~\bibnamefont {Ahlers}}, \bibinfo {author} {\bibfnamefont {S.}~\bibnamefont {Grossmann}},\ and\ \bibinfo {author} {\bibfnamefont {D.}~\bibnamefont {Lohse}},\ }\bibfield  {title} {\bibinfo {title} {{Heat transfer and large scale dynamics in turbulent Rayleigh-B{\'e}nard convection}},\ }\href@noop {} {\bibfield  {journal} {\bibinfo  {journal} {Rev. Mod. Phys.}\ }\textbf {\bibinfo {volume} {81}},\ \bibinfo {pages} {503} (\bibinfo {year} {2009})}\BibitemShut {NoStop}%
\bibitem [{\citenamefont {Chill{\`a}}\ and\ \citenamefont {Schumacher}(2012)}]{Chilla:EPJE2012}%
  \BibitemOpen
  \bibfield  {author} {\bibinfo {author} {\bibfnamefont {F.}~\bibnamefont {Chill{\`a}}}\ and\ \bibinfo {author} {\bibfnamefont {J.}~\bibnamefont {Schumacher}},\ }\bibfield  {title} {\bibinfo {title} {{New perspectives in turbulent Rayleigh-B{\'e}nard convection}},\ }\href@noop {} {\bibfield  {journal} {\bibinfo  {journal} {Eur. Phys. J. E}\ }\textbf {\bibinfo {volume} {35}},\ \bibinfo {pages} {58} (\bibinfo {year} {2012})}\BibitemShut {NoStop}%
\bibitem [{\citenamefont {Verma}(2018)}]{Verma:book:BDF}%
  \BibitemOpen
  \bibfield  {author} {\bibinfo {author} {\bibfnamefont {M.~K.}\ \bibnamefont {Verma}},\ }\href@noop {} {\emph {\bibinfo {title} {Physics of Buoyant Flows: From Instabilities to Turbulence}}}\ (\bibinfo  {publisher} {World Scientific},\ \bibinfo {address} {Singapore},\ \bibinfo {year} {2018})\BibitemShut {NoStop}%
\bibitem [{\citenamefont {Schumacher}\ and\ \citenamefont {Sreenivasan}(2020)}]{Schumacher:RMP2020}%
  \BibitemOpen
  \bibfield  {author} {\bibinfo {author} {\bibfnamefont {J.}~\bibnamefont {Schumacher}}\ and\ \bibinfo {author} {\bibfnamefont {K.~R.}\ \bibnamefont {Sreenivasan}},\ }\bibfield  {title} {\bibinfo {title} {Colloquium: Unusual dynamics of convection in the {S}un},\ }\href@noop {} {\bibfield  {journal} {\bibinfo  {journal} {Rev. Mod. Phys.}\ }\textbf {\bibinfo {volume} {92}},\ \bibinfo {pages} {041001} (\bibinfo {year} {2020})}\BibitemShut {NoStop}%
\bibitem [{\citenamefont {Spruit}\ \emph {et~al.}(1990)\citenamefont {Spruit}, \citenamefont {Nordlund},\ and\ \citenamefont {Title}}]{Spruit:ARAA1990}%
  \BibitemOpen
  \bibfield  {author} {\bibinfo {author} {\bibfnamefont {H.~C.}\ \bibnamefont {Spruit}}, \bibinfo {author} {\bibfnamefont {A.}~\bibnamefont {Nordlund}},\ and\ \bibinfo {author} {\bibfnamefont {A.~M.}\ \bibnamefont {Title}},\ }\bibfield  {title} {\bibinfo {title} {{Solar convection}},\ }\href@noop {} {\bibfield  {journal} {\bibinfo  {journal} {Annu. Rev. Astron. Astrophys.}\ }\textbf {\bibinfo {volume} {28}},\ \bibinfo {pages} {263} (\bibinfo {year} {1990})}\BibitemShut {NoStop}%
\bibitem [{\citenamefont {Shraiman}\ and\ \citenamefont {Siggia}(1990)}]{Shraiman:PRA1990}%
  \BibitemOpen
  \bibfield  {author} {\bibinfo {author} {\bibfnamefont {B.~I.}\ \bibnamefont {Shraiman}}\ and\ \bibinfo {author} {\bibfnamefont {E.~D.}\ \bibnamefont {Siggia}},\ }\bibfield  {title} {\bibinfo {title} {{Heat transport in high-Rayleigh-number convection}},\ }\href@noop {} {\bibfield  {journal} {\bibinfo  {journal} {Phys. Rev. A}\ }\textbf {\bibinfo {volume} {42}},\ \bibinfo {pages} {3650} (\bibinfo {year} {1990})}\BibitemShut {NoStop}%
\bibitem [{\citenamefont {Kraichnan}(1962)}]{Kraichnan:PF1962Convection}%
  \BibitemOpen
  \bibfield  {author} {\bibinfo {author} {\bibfnamefont {R.~H.}\ \bibnamefont {Kraichnan}},\ }\bibfield  {title} {\bibinfo {title} {{Turbulent thermal convection at arbitrary Prandtl number}},\ }\href@noop {} {\bibfield  {journal} {\bibinfo  {journal} {Phys. Fluids}\ }\textbf {\bibinfo {volume} {5}},\ \bibinfo {pages} {1374} (\bibinfo {year} {1962})}\BibitemShut {NoStop}%
\bibitem [{\citenamefont {Malkus}(1954)}]{Malkus:PRSA1954}%
  \BibitemOpen
  \bibfield  {author} {\bibinfo {author} {\bibfnamefont {W.~V.~R.}\ \bibnamefont {Malkus}},\ }\bibfield  {title} {\bibinfo {title} {{The Heat Transport and Spectrum of Thermal Turbulence}},\ }\href@noop {} {\bibfield  {journal} {\bibinfo  {journal} {Proc. R. Soc. Lond. A}\ }\textbf {\bibinfo {volume} {225}},\ \bibinfo {pages} {196} (\bibinfo {year} {1954})}\BibitemShut {NoStop}%
\bibitem [{\citenamefont {Grossmann}\ and\ \citenamefont {Lohse}(2000)}]{Grossmann:JFM2000}%
  \BibitemOpen
  \bibfield  {author} {\bibinfo {author} {\bibfnamefont {S.}~\bibnamefont {Grossmann}}\ and\ \bibinfo {author} {\bibfnamefont {D.}~\bibnamefont {Lohse}},\ }\bibfield  {title} {\bibinfo {title} {{Scaling in thermal convection: a unifying theory}},\ }\href@noop {} {\bibfield  {journal} {\bibinfo  {journal} {J. Fluid Mech.}\ }\textbf {\bibinfo {volume} {407}},\ \bibinfo {pages} {27} (\bibinfo {year} {2000})}\BibitemShut {NoStop}%
\bibitem [{\citenamefont {Chavanne}\ \emph {et~al.}(2001)\citenamefont {Chavanne}, \citenamefont {Chill{\`a}}, \citenamefont {Chabaud}, \citenamefont {Castaing},\ and\ \citenamefont {Hebral}}]{Chavanne:PF2001}%
  \BibitemOpen
  \bibfield  {author} {\bibinfo {author} {\bibfnamefont {X.}~\bibnamefont {Chavanne}}, \bibinfo {author} {\bibfnamefont {F.}~\bibnamefont {Chill{\`a}}}, \bibinfo {author} {\bibfnamefont {B.}~\bibnamefont {Chabaud}}, \bibinfo {author} {\bibfnamefont {B.}~\bibnamefont {Castaing}},\ and\ \bibinfo {author} {\bibfnamefont {B.}~\bibnamefont {Hebral}},\ }\bibfield  {title} {\bibinfo {title} {{Turbulent Rayleigh-B{\'e}nard convection in gaseous and liquid He}},\ }\href@noop {} {\bibfield  {journal} {\bibinfo  {journal} {Phys. Fluids}\ }\textbf {\bibinfo {volume} {13}},\ \bibinfo {pages} {1300} (\bibinfo {year} {2001})}\BibitemShut {NoStop}%
\bibitem [{\citenamefont {He}\ \emph {et~al.}(2012)\citenamefont {He}, \citenamefont {Funfschilling}, \citenamefont {Nobach}, \citenamefont {Bodenschatz},\ and\ \citenamefont {Ahlers}}]{He:PRL2012}%
  \BibitemOpen
  \bibfield  {author} {\bibinfo {author} {\bibfnamefont {X.}~\bibnamefont {He}}, \bibinfo {author} {\bibfnamefont {D.}~\bibnamefont {Funfschilling}}, \bibinfo {author} {\bibfnamefont {H.}~\bibnamefont {Nobach}}, \bibinfo {author} {\bibfnamefont {E.}~\bibnamefont {Bodenschatz}},\ and\ \bibinfo {author} {\bibfnamefont {G.}~\bibnamefont {Ahlers}},\ }\bibfield  {title} {\bibinfo {title} {{Transition to the Ultimate State of Turbulent Rayleigh-B{\'e}nard Convection}},\ }\href@noop {} {\bibfield  {journal} {\bibinfo  {journal} {Phys. Rev. Lett.}\ }\textbf {\bibinfo {volume} {108}},\ \bibinfo {pages} {024502} (\bibinfo {year} {2012})}\BibitemShut {NoStop}%
\bibitem [{\citenamefont {Zhu}\ \emph {et~al.}(2018)\citenamefont {Zhu}, \citenamefont {Mathai}, \citenamefont {Stevens}, \citenamefont {Verzicco},\ and\ \citenamefont {Lohse}}]{Zhu:PRL2018}%
  \BibitemOpen
  \bibfield  {author} {\bibinfo {author} {\bibfnamefont {X.}~\bibnamefont {Zhu}}, \bibinfo {author} {\bibfnamefont {V.}~\bibnamefont {Mathai}}, \bibinfo {author} {\bibfnamefont {R.~J. A.~M.}\ \bibnamefont {Stevens}}, \bibinfo {author} {\bibfnamefont {R.}~\bibnamefont {Verzicco}},\ and\ \bibinfo {author} {\bibfnamefont {D.}~\bibnamefont {Lohse}},\ }\bibfield  {title} {\bibinfo {title} {{Transition to the Ultimate Regime in Two-Dimensional Rayleigh-B{\'e}nard Convection}},\ }\href@noop {} {\bibfield  {journal} {\bibinfo  {journal} {Phys. Rev. Lett.}\ }\textbf {\bibinfo {volume} {120}},\ \bibinfo {pages} {144502} (\bibinfo {year} {2018})}\BibitemShut {NoStop}%
\bibitem [{\citenamefont {Lohse}\ and\ \citenamefont {Shishkina}(2024)}]{Lohse:RMP2024}%
  \BibitemOpen
  \bibfield  {author} {\bibinfo {author} {\bibfnamefont {D.}~\bibnamefont {Lohse}}\ and\ \bibinfo {author} {\bibfnamefont {O.}~\bibnamefont {Shishkina}},\ }\bibfield  {title} {\bibinfo {title} {{Ultimate Rayleigh-B{\'e}nard turbulence}},\ }\href@noop {} {\bibfield  {journal} {\bibinfo  {journal} {Rev. Mod. Phys.}\ }\textbf {\bibinfo {volume} {96}},\ \bibinfo {pages} {035001} (\bibinfo {year} {2024})}\BibitemShut {NoStop}%
\bibitem [{\citenamefont {Shishkina}\ and\ \citenamefont {Lohse}(2024)}]{Shishkina:PRL2024}%
  \BibitemOpen
  \bibfield  {author} {\bibinfo {author} {\bibfnamefont {O.}~\bibnamefont {Shishkina}}\ and\ \bibinfo {author} {\bibfnamefont {D.}~\bibnamefont {Lohse}},\ }\bibfield  {title} {\bibinfo {title} {Ultimate regime of rayleigh-b\'enard turbulence: Subregimes and their scaling relations for the nusselt vs rayleigh and prandtl numbers},\ }\href {https://doi.org/10.1103/PhysRevLett.133.144001} {\bibfield  {journal} {\bibinfo  {journal} {Phys. Rev. Lett.}\ }\textbf {\bibinfo {volume} {133}},\ \bibinfo {pages} {144001} (\bibinfo {year} {2024})}\BibitemShut {NoStop}%
\bibitem [{\citenamefont {Niemela}\ \emph {et~al.}(2000)\citenamefont {Niemela}, \citenamefont {Skrbek}, \citenamefont {Sreenivasan},\ and\ \citenamefont {Donnelly}}]{Niemela:Nature2000}%
  \BibitemOpen
  \bibfield  {author} {\bibinfo {author} {\bibfnamefont {J.~J.}\ \bibnamefont {Niemela}}, \bibinfo {author} {\bibfnamefont {L.}~\bibnamefont {Skrbek}}, \bibinfo {author} {\bibfnamefont {K.~R.}\ \bibnamefont {Sreenivasan}},\ and\ \bibinfo {author} {\bibfnamefont {R.~J.}\ \bibnamefont {Donnelly}},\ }\bibfield  {title} {\bibinfo {title} {{Turbulent convection at very high Rayleigh numbers}},\ }\href@noop {} {\bibfield  {journal} {\bibinfo  {journal} {Nature}\ }\textbf {\bibinfo {volume} {404}},\ \bibinfo {pages} {837} (\bibinfo {year} {2000})}\BibitemShut {NoStop}%
\bibitem [{\citenamefont {Urban}\ \emph {et~al.}(2012)\citenamefont {Urban}, \citenamefont {Hanzelka}, \citenamefont {Kralik}, \citenamefont {Musilov{\'a}}, \citenamefont {Srnka},\ and\ \citenamefont {Skrbek}}]{Urban:PRL2012}%
  \BibitemOpen
  \bibfield  {author} {\bibinfo {author} {\bibfnamefont {P.}~\bibnamefont {Urban}}, \bibinfo {author} {\bibfnamefont {P.}~\bibnamefont {Hanzelka}}, \bibinfo {author} {\bibfnamefont {T.}~\bibnamefont {Kralik}}, \bibinfo {author} {\bibfnamefont {V.}~\bibnamefont {Musilov{\'a}}}, \bibinfo {author} {\bibfnamefont {A.}~\bibnamefont {Srnka}},\ and\ \bibinfo {author} {\bibfnamefont {L.}~\bibnamefont {Skrbek}},\ }\bibfield  {title} {\bibinfo {title} {{Effect of Boundary Layers Asymmetry on Heat Transfer Efficiency in Turbulent Rayleigh-B{\'e}nard Convection at Very High Rayleigh Numbers}},\ }\href@noop {} {\bibfield  {journal} {\bibinfo  {journal} {Phys. Rev. Lett.}\ }\textbf {\bibinfo {volume} {109}},\ \bibinfo {pages} {154301} (\bibinfo {year} {2012})}\BibitemShut {NoStop}%
\bibitem [{\citenamefont {Iyer}\ \emph {et~al.}(2020)\citenamefont {Iyer}, \citenamefont {Scheel}, \citenamefont {Schumacher},\ and\ \citenamefont {Sreenivasan}}]{Iyer:PNAS2020}%
  \BibitemOpen
  \bibfield  {author} {\bibinfo {author} {\bibfnamefont {K.~P.}\ \bibnamefont {Iyer}}, \bibinfo {author} {\bibfnamefont {J.~D.}\ \bibnamefont {Scheel}}, \bibinfo {author} {\bibfnamefont {J.}~\bibnamefont {Schumacher}},\ and\ \bibinfo {author} {\bibfnamefont {K.~R.}\ \bibnamefont {Sreenivasan}},\ }\bibfield  {title} {\bibinfo {title} {{Classical 1/3 scaling of convection holds up to Ra$~= 10^{15}$}},\ }\href@noop {} {\bibfield  {journal} {\bibinfo  {journal} {Proc. Natl. Acad. Sci.}\ }\textbf {\bibinfo {volume} {117}},\ \bibinfo {pages} {7594} (\bibinfo {year} {2020})}\BibitemShut {NoStop}%
\bibitem [{\citenamefont {Ahlers}\ \emph {et~al.}(2012)\citenamefont {Ahlers}, \citenamefont {Bodenschatz}, \citenamefont {Funfschilling}, \citenamefont {Grossmann}, \citenamefont {He}, \citenamefont {Lohse}, \citenamefont {Stevens},\ and\ \citenamefont {Verzicco}}]{Ahlers:PRL2012}%
  \BibitemOpen
  \bibfield  {author} {\bibinfo {author} {\bibfnamefont {G.}~\bibnamefont {Ahlers}}, \bibinfo {author} {\bibfnamefont {E.}~\bibnamefont {Bodenschatz}}, \bibinfo {author} {\bibfnamefont {D.}~\bibnamefont {Funfschilling}}, \bibinfo {author} {\bibfnamefont {S.}~\bibnamefont {Grossmann}}, \bibinfo {author} {\bibfnamefont {X.}~\bibnamefont {He}}, \bibinfo {author} {\bibfnamefont {D.}~\bibnamefont {Lohse}}, \bibinfo {author} {\bibfnamefont {R.~J. A.~M.}\ \bibnamefont {Stevens}},\ and\ \bibinfo {author} {\bibfnamefont {R.}~\bibnamefont {Verzicco}},\ }\bibfield  {title} {\bibinfo {title} {{Logarithmic Temperature Profiles in Turbulent Rayleigh-B{\'e}nard Convection}},\ }\href@noop {} {\bibfield  {journal} {\bibinfo  {journal} {Phys. Rev. Lett.}\ }\textbf {\bibinfo {volume} {109}},\ \bibinfo {pages} {114501} (\bibinfo {year} {2012})}\BibitemShut {NoStop}%
\bibitem [{\citenamefont {Urban}\ \emph {et~al.}(2011)\citenamefont {Urban}, \citenamefont {Musilov{\'a}},\ and\ \citenamefont {Skrbek}}]{Urban:PRL2011}%
  \BibitemOpen
  \bibfield  {author} {\bibinfo {author} {\bibfnamefont {P.}~\bibnamefont {Urban}}, \bibinfo {author} {\bibfnamefont {V.}~\bibnamefont {Musilov{\'a}}},\ and\ \bibinfo {author} {\bibfnamefont {L.}~\bibnamefont {Skrbek}},\ }\bibfield  {title} {\bibinfo {title} {{Efficiency of heat transfer in turbulent Rayleigh-B{\'e}nard convection}},\ }\href@noop {} {\bibfield  {journal} {\bibinfo  {journal} {Phys. Rev. Lett.}\ }\textbf {\bibinfo {volume} {107}},\ \bibinfo {pages} {014302} (\bibinfo {year} {2011})}\BibitemShut {NoStop}%
\bibitem [{\citenamefont {Verma}\ \emph {et~al.}(2012)\citenamefont {Verma}, \citenamefont {Mishra}, \citenamefont {Pandey},\ and\ \citenamefont {Paul}}]{Verma:PRE2012}%
  \BibitemOpen
  \bibfield  {author} {\bibinfo {author} {\bibfnamefont {M.~K.}\ \bibnamefont {Verma}}, \bibinfo {author} {\bibfnamefont {P.~K.}\ \bibnamefont {Mishra}}, \bibinfo {author} {\bibfnamefont {A.}~\bibnamefont {Pandey}},\ and\ \bibinfo {author} {\bibfnamefont {S.}~\bibnamefont {Paul}},\ }\bibfield  {title} {\bibinfo {title} {{Scalings of field correlations and heat transport in turbulent convection}},\ }\href@noop {} {\bibfield  {journal} {\bibinfo  {journal} {Phys. Rev. E}\ }\textbf {\bibinfo {volume} {85}},\ \bibinfo {pages} {016310} (\bibinfo {year} {2012})}\BibitemShut {NoStop}%
\bibitem [{\citenamefont {John}\ and\ \citenamefont {Schumacher}(2023{\natexlab{a}})}]{John:PRF2023}%
  \BibitemOpen
  \bibfield  {author} {\bibinfo {author} {\bibfnamefont {J.~P.}\ \bibnamefont {John}}\ and\ \bibinfo {author} {\bibfnamefont {J.}~\bibnamefont {Schumacher}},\ }\bibfield  {title} {\bibinfo {title} {Strongly superadiabatic and stratified limits of compressible convection},\ }\href@noop {} {\bibfield  {journal} {\bibinfo  {journal} {Phys. Rev. Fluids}\ }\textbf {\bibinfo {volume} {8}},\ \bibinfo {pages} {103505} (\bibinfo {year} {2023}{\natexlab{a}})}\BibitemShut {NoStop}%
\bibitem [{\citenamefont {John}\ and\ \citenamefont {Schumacher}(2023{\natexlab{b}})}]{John:JFM2023}%
  \BibitemOpen
  \bibfield  {author} {\bibinfo {author} {\bibfnamefont {J.~P.}\ \bibnamefont {John}}\ and\ \bibinfo {author} {\bibfnamefont {J.}~\bibnamefont {Schumacher}},\ }\bibfield  {title} {\bibinfo {title} {Compressible turbulent convection in highly stratified adiabatic background},\ }\href@noop {} {\bibfield  {journal} {\bibinfo  {journal} {J. Fluid Mech.}\ }\textbf {\bibinfo {volume} {972}},\ \bibinfo {pages} {R4} (\bibinfo {year} {2023}{\natexlab{b}})}\BibitemShut {NoStop}%
\bibitem [{\citenamefont {Porter}\ and\ \citenamefont {Woodward}(2000)}]{Porter:AJS2000}%
  \BibitemOpen
  \bibfield  {author} {\bibinfo {author} {\bibfnamefont {D.~H.}\ \bibnamefont {Porter}}\ and\ \bibinfo {author} {\bibfnamefont {P.~R.}\ \bibnamefont {Woodward}},\ }\bibfield  {title} {\bibinfo {title} {Three-dimensional simulations of turbulent compressible convection},\ }\href@noop {} {\bibfield  {journal} {\bibinfo  {journal} {Astrophys. J. Suppl. Ser.}\ }\textbf {\bibinfo {volume} {127}},\ \bibinfo {pages} {159} (\bibinfo {year} {2000})}\BibitemShut {NoStop}%
\bibitem [{\citenamefont {Tiwari}\ \emph {et~al.}(2024)\citenamefont {Tiwari}, \citenamefont {Sharma},\ and\ \citenamefont {Verma}}]{Tiwari:arxiv2024}%
  \BibitemOpen
  \bibfield  {author} {\bibinfo {author} {\bibfnamefont {H.}~\bibnamefont {Tiwari}}, \bibinfo {author} {\bibfnamefont {L.}~\bibnamefont {Sharma}},\ and\ \bibinfo {author} {\bibfnamefont {M.~K.}\ \bibnamefont {Verma}},\ }\bibfield  {title} {\bibinfo {title} {Compressible turbulent convection at very high rayleigh numbers},\ }\href@noop {} {\  (\bibinfo {year} {2024})},\ \Eprint {https://arxiv.org/abs/2411.10372} {arXiv:2411.10372 [physics.flu-dyn]} \BibitemShut {NoStop}%
\bibitem [{\citenamefont {Ouyang}\ \emph {et~al.}(2013)\citenamefont {Ouyang}, \citenamefont {He}, \citenamefont {Xu}, \citenamefont {Luo},\ and\ \citenamefont {Zhang}}]{Ouyang:CG2013}%
  \BibitemOpen
  \bibfield  {author} {\bibinfo {author} {\bibfnamefont {C.}~\bibnamefont {Ouyang}}, \bibinfo {author} {\bibfnamefont {S.}~\bibnamefont {He}}, \bibinfo {author} {\bibfnamefont {Q.}~\bibnamefont {Xu}}, \bibinfo {author} {\bibfnamefont {Y.}~\bibnamefont {Luo}},\ and\ \bibinfo {author} {\bibfnamefont {W.}~\bibnamefont {Zhang}},\ }\bibfield  {title} {\bibinfo {title} {{A MacCormack-TVD finite difference method to simulate the mass flow in mountainous terrain with variable computational domain}},\ }\href@noop {} {\bibfield  {journal} {\bibinfo  {journal} {Comput. Geosci.}\ }\textbf {\bibinfo {volume} {52}},\ \bibinfo {pages} {1} (\bibinfo {year} {2013})}\BibitemShut {NoStop}%
\bibitem [{\citenamefont {Yee}(1987)}]{Yee:book}%
  \BibitemOpen
  \bibfield  {author} {\bibinfo {author} {\bibfnamefont {H.~C.}\ \bibnamefont {Yee}},\ }\href@noop {} {\emph {\bibinfo {title} {Upwind and symmetric shock-capturing schemes}}}\ (\bibinfo  {publisher} {National Aeronautics and Space Administration, Ames Research Center},\ \bibinfo {year} {1987})\BibitemShut {NoStop}%
\bibitem [{\citenamefont {Liang}\ \emph {et~al.}(2007)\citenamefont {Liang}, \citenamefont {Lin},\ and\ \citenamefont {Falconer}}]{Liang:IJNMF2007}%
  \BibitemOpen
  \bibfield  {author} {\bibinfo {author} {\bibfnamefont {D.}~\bibnamefont {Liang}}, \bibinfo {author} {\bibfnamefont {B.}~\bibnamefont {Lin}},\ and\ \bibinfo {author} {\bibfnamefont {R.~A.}\ \bibnamefont {Falconer}},\ }\bibfield  {title} {\bibinfo {title} {{Simulation of rapidly varying flow using an efficient TVD-MacCormack scheme}},\ }\href@noop {} {\bibfield  {journal} {\bibinfo  {journal} {Int. J. Numer. Methods Fluids}\ }\textbf {\bibinfo {volume} {53}},\ \bibinfo {pages} {811} (\bibinfo {year} {2007})}\BibitemShut {NoStop}%
\bibitem [{\citenamefont {{Wesseling}}(2000)}]{Wesseling:book:CFD}%
  \BibitemOpen
  \bibfield  {author} {\bibinfo {author} {\bibnamefont {{Wesseling}}},\ }\href@noop {} {\emph {\bibinfo {title} {{Principles of Computational Fluid Dynamics}}}},\ \bibinfo {edition} {1st}\ ed.\ (\bibinfo  {publisher} {Springer-Verlag},\ \bibinfo {address} {Berlin Heidelberg},\ \bibinfo {year} {2000})\BibitemShut {NoStop}%
\bibitem [{\citenamefont {Pandey}\ and\ \citenamefont {Verma}(2016)}]{Pandey:PF2016}%
  \BibitemOpen
  \bibfield  {author} {\bibinfo {author} {\bibfnamefont {A.}~\bibnamefont {Pandey}}\ and\ \bibinfo {author} {\bibfnamefont {M.~K.}\ \bibnamefont {Verma}},\ }\bibfield  {title} {\bibinfo {title} {{Scaling of large-scale quantities in Rayleigh-B{\'e}nard convection}},\ }\href@noop {} {\bibfield  {journal} {\bibinfo  {journal} {Phys. Fluids}\ }\textbf {\bibinfo {volume} {28}},\ \bibinfo {pages} {095105} (\bibinfo {year} {2016})}\BibitemShut {NoStop}%
\bibitem [{\citenamefont {Spiegel}(1965)}]{Spiegel:AJ1965}%
  \BibitemOpen
  \bibfield  {author} {\bibinfo {author} {\bibfnamefont {E.~A.}\ \bibnamefont {Spiegel}},\ }\bibfield  {title} {\bibinfo {title} {{Convective instability in a compressible atmosphere. I.}},\ }\href@noop {} {\bibfield  {journal} {\bibinfo  {journal} {Astrophys. J.}\ }\textbf {\bibinfo {volume} {141}},\ \bibinfo {pages} {1068} (\bibinfo {year} {1965})}\BibitemShut {NoStop}%
\bibitem [{\citenamefont {Graham}(1975)}]{Graham:JFM1975}%
  \BibitemOpen
  \bibfield  {author} {\bibinfo {author} {\bibfnamefont {E.}~\bibnamefont {Graham}},\ }\bibfield  {title} {\bibinfo {title} {Numerical simulation of two-dimensional compressible convection},\ }\href@noop {} {\bibfield  {journal} {\bibinfo  {journal} {J. Fluid Mech.}\ }\textbf {\bibinfo {volume} {70}},\ \bibinfo {pages} {689} (\bibinfo {year} {1975})}\BibitemShut {NoStop}%
\bibitem [{\citenamefont {Jagannathan}\ and\ \citenamefont {Donzis}(2016)}]{Jagannathan:JFM2016}%
  \BibitemOpen
  \bibfield  {author} {\bibinfo {author} {\bibfnamefont {S.}~\bibnamefont {Jagannathan}}\ and\ \bibinfo {author} {\bibfnamefont {D.~A.}\ \bibnamefont {Donzis}},\ }\bibfield  {title} {\bibinfo {title} {Reynolds and mach number scaling in solenoidally-forced compressible turbulence using high-resolution direct numerical simulations},\ }\href {https://doi.org/10.1017/jfm.2015.754} {\bibfield  {journal} {\bibinfo  {journal} {Journal of Fluid Mechanics}\ }\textbf {\bibinfo {volume} {789}},\ \bibinfo {pages} {669–707} (\bibinfo {year} {2016})}\BibitemShut {NoStop}%
\bibitem [{\citenamefont {Gr{\"o}tzbach}(1983)}]{Grotzbach:JCP1983}%
  \BibitemOpen
  \bibfield  {author} {\bibinfo {author} {\bibfnamefont {G.}~\bibnamefont {Gr{\"o}tzbach}},\ }\bibfield  {title} {\bibinfo {title} {Spatial resolution requirements for direct numerical simulation of the {R}ayleigh-{B}{\'e}nard convection},\ }\href@noop {} {\bibfield  {journal} {\bibinfo  {journal} {J. Comput. Phys.}\ }\textbf {\bibinfo {volume} {49}},\ \bibinfo {pages} {241} (\bibinfo {year} {1983})}\BibitemShut {NoStop}%
\bibitem [{\citenamefont {Landau}\ and\ \citenamefont {Lifshitz}(1987)}]{Landau:book:Fluid}%
  \BibitemOpen
  \bibfield  {author} {\bibinfo {author} {\bibfnamefont {L.~D.}\ \bibnamefont {Landau}}\ and\ \bibinfo {author} {\bibfnamefont {E.~M.}\ \bibnamefont {Lifshitz}},\ }\href@noop {} {\emph {\bibinfo {title} {{Fluid Mechanics}}}},\ \bibinfo {edition} {2nd}\ ed.,\ Course of Theoretical Physics\ (\bibinfo  {publisher} {Elsevier},\ \bibinfo {address} {Oxford},\ \bibinfo {year} {1987})\BibitemShut {NoStop}%
\bibitem [{\citenamefont {Verhoeven}\ \emph {et~al.}(2015)\citenamefont {Verhoeven}, \citenamefont {Wieseh{\"o}fer},\ and\ \citenamefont {Stellmach}}]{Verhoeven:AJ2015}%
  \BibitemOpen
  \bibfield  {author} {\bibinfo {author} {\bibfnamefont {J.}~\bibnamefont {Verhoeven}}, \bibinfo {author} {\bibfnamefont {T.}~\bibnamefont {Wieseh{\"o}fer}},\ and\ \bibinfo {author} {\bibfnamefont {S.}~\bibnamefont {Stellmach}},\ }\bibfield  {title} {\bibinfo {title} {Anelastic versus fully compressible turbulent {R}ayleigh--{B}{\'e}nard convection},\ }\href@noop {} {\bibfield  {journal} {\bibinfo  {journal} {Astrophys. J.}\ }\textbf {\bibinfo {volume} {805}},\ \bibinfo {pages} {62} (\bibinfo {year} {2015})}\BibitemShut {NoStop}%
\bibitem [{\citenamefont {Parker}(1970)}]{Parker:ARAA1970}%
  \BibitemOpen
  \bibfield  {author} {\bibinfo {author} {\bibfnamefont {E.~N.}\ \bibnamefont {Parker}},\ }\bibfield  {title} {\bibinfo {title} {{The origin of solar magnetic fields}},\ }\href@noop {} {\bibfield  {journal} {\bibinfo  {journal} {Annu. Rev. Astron. Astrophys.}\ }\textbf {\bibinfo {volume} {8}},\ \bibinfo {pages} {1} (\bibinfo {year} {1970})}\BibitemShut {NoStop}%
\bibitem [{\citenamefont {Miesch}(2005)}]{Miesch:LRSP2005}%
  \BibitemOpen
  \bibfield  {author} {\bibinfo {author} {\bibfnamefont {M.~S.}\ \bibnamefont {Miesch}},\ }\bibfield  {title} {\bibinfo {title} {Large-scale dynamics of the convection zone and tachocline},\ }\href@noop {} {\bibfield  {journal} {\bibinfo  {journal} {Living Reviews in Solar Physics}\ }\textbf {\bibinfo {volume} {2}},\ \bibinfo {pages} {1} (\bibinfo {year} {2005})}\BibitemShut {NoStop}%
\end{thebibliography}%

\end{document}